\documentclass[iop]{emulateapj}
\slugcomment{to be submitted to {\it The Astrophysical Journal}}
\shortauthors{Istrate et al.}

\newcommand{\gta}{\lower 0.5ex\hbox{$ \buildrel>\over\sim\ $}}
\newcommand{\lta}{\lower 0.5ex\hbox{$ \buildrel<\over\sim\ $}}

\newcommand{\teff}{$T_{\rm eff}$}
\newcommand{\Msun}{$\mathrm{M}_{\odot}$}

\usepackage{natbib}
\usepackage{graphicx}
\usepackage[backref=blue,breaklinks,colorlinks,citecolor=blue]{hyperref}
\begin{document}

\title{A Model of the Pulsating Extremely Low-Mass White Dwarf Precursor
  WASP 0247$-$25B}

\author{A.G. Istrate\altaffilmark{1}, G. Fontaine\altaffilmark{2},
  and C. Heuser\altaffilmark{3}}

\altaffiltext{1}{Center for Gravitation, Cosmology, and Astrophysics,
  Department of Physics, University of Wisconsin-Milwaukee, P.O. Box
  413, Milwaukee, WI 53201,USA; istrate@uwm.edu}
\altaffiltext{2}{D\'epartement de Physique, Universit\'e de Montr\'eal,
  C.P. 6128, Succursale Centre-Ville, Montr\'eal, QC H3C 3J7, Canada} 
\altaffiltext{3}{Dr. Karl Remeis-Observatory \& ECAP, Astronomical
  Institute, Friedrich-Alexander University Erlangen-N\"urnberg,
  Sternwartstr. 7, 96049 Bamberg, Germany}

\begin{abstract}

We present an analysis of the evolutionary and pulsation properties of
the extremely low-mass white dwarf precursor (B) component of the
double-lined eclipsing system WASP 0247$-$25. Given that the fundamental
parameters of that star have been obtained previously at a unique level
of precision, WASP 0247$-$25B represents the ideal case for testing evolutionary models of this newly-found category of pulsators. 
Taking into account the known constraints on the mass, orbital period,
effective temperature, surface gravity, and atmospheric composition, we
present a model that is compatible with these constraints and show
pulsation modes that have periods very close to the observed values. 
Importantly, these modes are predicted excited. Although the overall consistency
remains perfectible, the observable properties of WASP 0247$-$25B are
closely reproduced. A key ingredient of our binary evolutionary models is represented by rotational mixing as the main competitor against gravitational settling. Depending on assumptions made about the values
of the degree index $\ell$ for the observed pulsation modes, we found three
possible seismic solutions. We discuss two tests, rotational splitting
and multicolor photometry, that should readily identify the modes and
discriminate between these solutions. However, this will require improved
temporal resolution and higher S/N observations than currently
available. 

\end{abstract}


\section{Introduction}

The discovery of multi-periodic pulsations in the extremely low-mass white dwarf precursor (ELM proto-WD) WASP 0247$-$25B by \cite{maxted2013} has opened the way for the application of asteroseismic methods for testing and constraining evolutionary  models of these intriguing stars believed to descend from stripped red giants through binary evolution. This was followed a year later by the report of \cite{maxted2014a} on luminosity variations with comparable periods coming again from the ELM proto-WD component in another close binary system, WASP 1628+10, belonging also to the so-called EL CVn type.  The EL CVn binaries are rare double-lined eclipsing systems characterized by an A-type primary that outshines the smaller but hotter ELM proto-WD secondary, except in the UV (see, e.g., 
\citealt{maxted2014b}).  The analyses of the eclipses carried out by \citet[]{maxted2013, maxted2014a}, combining multicolor photometry and time-phased spectroscopy, have led to the determination of the fundamental stellar properties of both components of both systems at an impressive level of precision. In particular, in the stronger case of WASP 0247$-$25B, the very precise estimates of the fundamental parameters can be used as extremely useful data for comparing with evolutionary and seismic models that can currently be computed for ELM proto-WDs.

Multi-periodic pulsations with periods comparable to those found in WASP 0247$-$25B and WASP 1628+10B have also been discovered recently in three other ELM proto-WDs \citep{gianninas2016}\footnote{We disregard here   the two cool ELM proto-WD candidates proposed by \cite{corti2016}, and also the system discussed by  \cite{zhang2016}, as their nature is currently unclear \citep[see also the discussion in][]{brown2017}.}. Those belong to more frequent single-lined binary systems consisting of an essentially invisible primary, presumed to be a normal CO-core WD,  and of an ELM proto-WD as the secondary. Analyses of the optical spectra of these systems, whose light is thus totally dominated by the secondary, showed that the three new pulsators belong to the same region of the effective temperature-surface gravity plane as the two WASP stars discussed  above \citep{gianninas2014a,gianninas2015}. Additionally, it has been shown that the atmospheres of these stars contain amounts of helium comparable to those of hydrogen. 

 It is known that He is the essential ingredient, through He II-He III ionization, for driving pulsation modes in the regime of effective temperature where the known pulsating ELM proto-WDs are found, i.e., in a region where H is totally ionized and cannot contribute to pulsational driving \citep{jeffery2013, van_grootel2015, corsico2016, gianninas2016}. In this context, \cite{gianninas2016} provided the first empirical evidence that pulsations in ELM proto-WDs can only occur if a significant amount of He is present in the atmosphere and, by extension, in the driving region.   We note, in this respect, that the atmospheric He content has not been determined for the two WASP systems studied by \citet[]{maxted2013, maxted2014a}, given the complicated nature of their double-lined spectra (but see below).

Along with He, the atmospheres of ELM proto-WDs in general often show traces of metals such as Mg and Ca, as revealed in the systematic investigation of the class properties carried out by \cite{gianninas2014a}, and in the few available detailed abundance analyses of individual objects that currently exist \citep{kaplan2013, gianninas2014b, hermes2014, latour2016}. Given that gravitational settling, if left unimpeded, leads to the formation of a pure H atmosphere on a very short timescale in such stars \citep{althaus2001, althaus2013}, a  competing mechanism must be at work to account for the presence of elements heavier than hydrogen. Among the possible processes discussed by \cite{gianninas2014b}, radiative levitation alone certainly cannot
explain the relatively large amounts of He detected in ELM proto-WDs and, at least in one case, was shown to be unable to account for the observed pattern of metals \citep{hermes2014}. In that context, the last authors suggested that rotational mixing could be a mechanism of importance, given that it should occur naturally in close binary systems in which ELM proto-WDs are found. Moreover, \cite{corsico2016}  showed that in order to explain the existence of the ELM proto-WD pulsator class,  gravitational settling must be  counteracted by another process.
 
\cite{istrate2016a} were the first to investigate the combined effects of rotational mixing and diffusion processes (gravitational
settling, thermal diffusion, and chemical diffusion) on the (proto-) WDs that are formed through stable Roche--lobe overflow (the low-mass X-ray binary channel). They demonstrated that rotational mixing is indeed able to compete efficiently against gravitational settling and to account qualitatively for the presence of He and of metals (best represented by Ca in their calculations) in the atmospheres of ELM proto-WDs. After the end of the mass-transfer phase, the envelope of the newly formed proto-WD contracts significantly, rotating thus faster than the He core.  The differential rotation present in this evolutionary stage gives rise to rotational instabilities, with Eddington-Sweet currents being the main process responsible for the mixing of material, especially in the envelope (for more details, see \citet{istrate2016a}). Hence, rotational mixing appears as a natural mechanism within the framework of the low-mass X-ray binary scenario.

In a subsequent effort, \cite{istrate2016b} exploited the available seismic information on pulsating ELM proto-WDs to test the proposition that rotational mixing is likely a fundamental process in the formation and evolution of low-mass He-core white dwarfs. Specifically, they investigated whether rotational mixing can maintain a sufficient amount of He in the driving region of the star (deep in the envelope), such that it can fuel, through He II-He III ionization, the  observed pulsations in this type of stars. The test was successful, as in the region of the effective temperature-surface gravity domain where these pulsators are found, in a well-defined instability strip with blue and red edges, their models developed pulsational instabilities with characteristic periods that generally agree well with the detected periods. In addition, by the time such a model leaves the proto-WD instability strip, enters its cooling track, and crosses the blue edge of the ZZ Ceti instability strip,
gravitational settling has won over rotational mixing and the star has developed a pure H envelope. At this evolutionary stage, the computed model   is then able to drive pulsations again, but, this time, through pure H ionization (convective driving) and as a cool, low-mass ZZ Ceti star \citep{steinfadt2010, corsico2012, vangrootel2013}. This provides a natural explanation and a common origin for both types of pulsating ELM WDs  \citep{fontaine2017}. We note that these two families of pulsators are related to the classical pulsating white dwarfs whose properties have been reviewed by \cite{winget2008}, \cite{fontaine2008} and \cite{althaus2010}.

In the present work, we consider a more stringent and quantitative test of the rotationally-mixed models of \cite{istrate2016a}, by
investigating to what extent they can reproduce the fundamental astrophysical parameters that have been inferred for the best-studied of the known pulsating ELM proto-WDs, i.e., WASP 0247$-$25B, while being consistent with the available seismic data. We thus exploit here the unique data set provided by \cite{maxted2013} for this eclipsing system. 
% Table 1
\begin{deluxetable}{cccc}
\tablewidth{0pt}
%\rotate
\tabletypesize{\small}
\tablecaption{Available Constraints}
\tablehead{
\colhead{Quantity} &
\colhead{Value} &
\colhead{Method} &
\colhead{Source} 
 }
\startdata
$M$(B) & 0.186 $\pm$ 0.002 \Msun & a & 1 \\
$M$(A) & 1.356 $\pm$ 0.007 \Msun & a & 1 \\
$P_{\rm orbit}$ & 16.027908 $\pm$ 0.000010 h & a & 1 \\
\teff(B)   & 11,380 $\pm$ 400 K & a & 1 \\
log $g$(B) & 4.576 $\pm$ 0.011 cm s$^{-2}$ & a & 1 \\
\teff(B)   & 10,870 $\pm$ 230 K & b & 2 \\
log $g$(B) & 4.70 $\pm^{0.11}_{0.12}$ cm s$^{-2}$ & b & 2 \\
log He/H(B) & $-$0.87 $\pm^{0.19}_{0.27}$ & b & 2 \\
log Ca/H(B) & $-$6.50 $\pm$ 0.013 & b & 2 \\
log O/H(B)  & $-$4.40$\pm^{0.08}_{0.09}$     & b & 2 \\
$P_1$ & 380.95 $\pm$ 0.17 s & c & 1 \\
$P_2$ & 405.82 $\pm$ 0.61 s & c & 1 \\
$P_3$ & 420.64 $\pm$ 0.76 s & c & 1 \\
\enddata
\tablecomments{
(A) the primary A-type star component; 
(B) the secondary ELM proto-WD;
(a) Multicolor photometry and time-phased spectroscopy;
(b) Spectroscopic analysis of the disentangled time-averaged spectrum;
(c) Multicolor photometry.}  
\tablerefs{(1) Maxted et al. (2013); (2) Heuser et al. (2017).}
\label{table:observed_parameters}
\end{deluxetable}

\section{STRATEGY}

We seek binary evolutionary models that have component masses that agree with the obtained values of \cite{maxted2013}. Ideally, the evolutionary track describing the B component should pass through the inferred location of WASP 0247$-$25B in the effective temperature-surface gravity plane, and have a proto-WD age consistent with the age obtained by satisfying the orbital period constraint. The proto-WD age is defined as the time elapsed since the end of the mass--transfer phase  \citep[for more details, see][]{istrate2014b}. In addition, the model of the B component at that same point in the \teff-log $g$ diagram should show pulsations in its period spectrum that agree with the three observed periods. In this, we restrict the search to degree values of $\ell$ = 0, 1, and 2 on the basis of geometric arguments for mode visibility\footnote{We note in this context that \cite{maxted2013} have already demonstrated that the spacings   between the three observed periods in WASP 0247$-$25B are inconsistent   with the idea of modes with the same value of the degree index   $\ell$. At least modes with degree values of $\ell$ = 0 and 1, and  perhaps 2, must be involved; see their Fig. 3.}. In particular, the modes of interest in the model should be predicted excited, meaning that rotational mixing must have retained enough He in the envelope for pulsation driving to occur.

In this quest, we also exploit additional information about the atmospheric properties of WASP 0247$-$25B that can be obtained from the work of \cite{maxted2013}. In order to check on the values of the effective temperature of the two
components derived from their detailed eclipse analysis, the authors report -- in the supplementary online material associated with their paper -- that they went through the difficult exercise of disentangling the combined spectrum. Their  preliminary analysis of the spectrum of the B component indicates a value of \teff~= 10,300 $\pm$ 200 K for a fixed value of log $g$ = 4.61 and a fixed metallicity of $Z$ = 0.002. These numbers are to be compared with the values of \teff~=11,380 $\pm$ 400 K and log $g$ = 4.576 $\pm$ 0.011 based on their primary analysis of the eclipsing light curve, which the authors preferred and reported in Table 1 of their main paper \citep{maxted2013}. A more involved study of the disentangled spectrum of WASP 0247$-$25B was postponed to a later date.

 Recently, one of us (C.H.) carried out a self-consistent analysis of that spectrum using the suite
of tools available at Dr. Karl Remeis-Observatory. The details of that quantitative study will be reported elsewhere, but we use here the derived values of  \teff~= 10,870 $\pm$ 230 K and log~$g$~=4.70$\pm^{0.11}_{0.12}$, which, importantly in the present context, provide an independent (purely spectroscopic) estimate of the location of WASP 0247$-$25B in the \teff-log $g$ plane. The reported uncertainties are the combined statistical errors (which are smaller than 0.02 dex) and the assumed systematical errors on log~$g$~(0.1 dex) and \teff~(2$\%$). We note that the two independent sets of estimates are not in direct conflict as they overlap at the 1-$\sigma$ level in both \teff~and log $g$. In addition, this analysis provides estimates of, among others, the He, O, and Ca contents in the atmosphere of WASP 0247$-$25B, values that can ultimately be compared with predictions of our evolutionary models.
\begin{figure}
\includegraphics[width=\columnwidth]{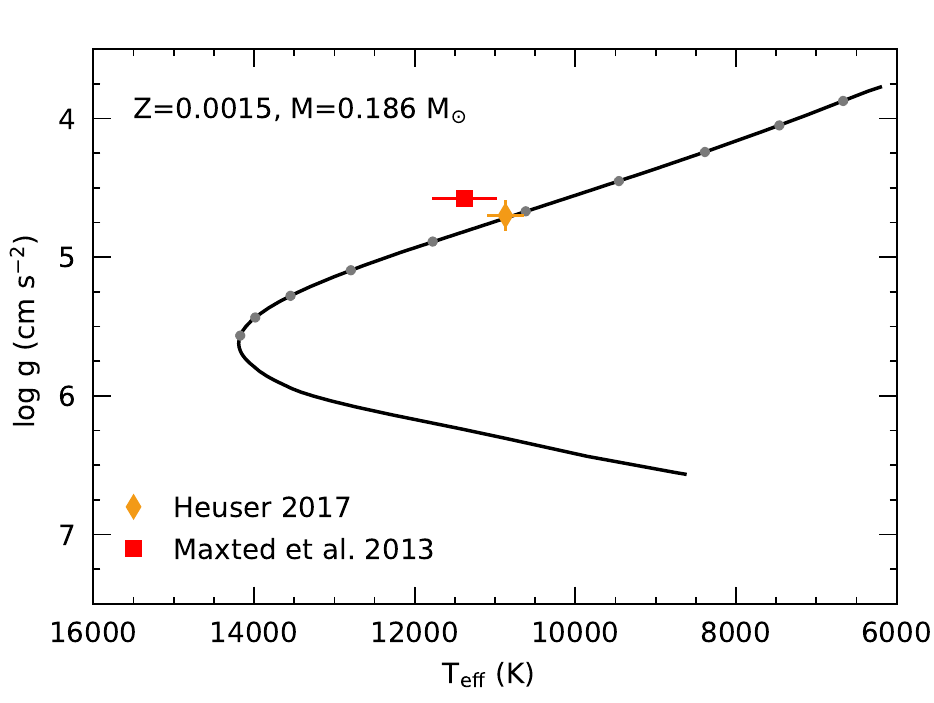}
\caption{ Evolutionary track  of our solution in the Kiel diagram from the beginning of the proto-WD phase until the system reached the age of 14 Gyr. The red square represents the atmospheric parameters as measured by  \cite{maxted2013}, while the orange diamond represent the values obtained by \cite{heuser2017}. The grey circles  indicate the proto-WD age in units of 100 Myr.}
\label{fig:fig1}
\vspace{0.1in}
\end{figure}

We have assembled, in Table~\ref{table:observed_parameters}, the constraints available in the pursuit of our exercise. Those include the two independent sets of \teff-log $g$ values discussed above, as well as the three detected periods in
WASP 0247$-$25B, $P_1$ through $P_3$. In a first step, we seek  evolutionary models that would best be compatible with the relevant entries listed in Table~\ref{table:observed_parameters}. We compute such models in a way similar to those described
in \cite{istrate2016a, istrate2016b}, using the publicly available binary stellar evolution code MESA, version 7624 \citep{paxton2011, paxton2013, paxton2015}, with rotational mixing and diffusion processes turned on. We specifically followed the evolution of H$^1$, He$^4$, C$^{12}$, O$^{16}$, and Ca$^{40}$ in the calculations presented here. We next carry out a detailed nonadiabatic pulsation study of these models  using the Montr\'eal pulsation code \citep{brassard1992,fontaine1994} in order to best fit the observed pulsation periods and verify if these modes are predicted excited. 

\begin{figure}
\includegraphics[width=\columnwidth]{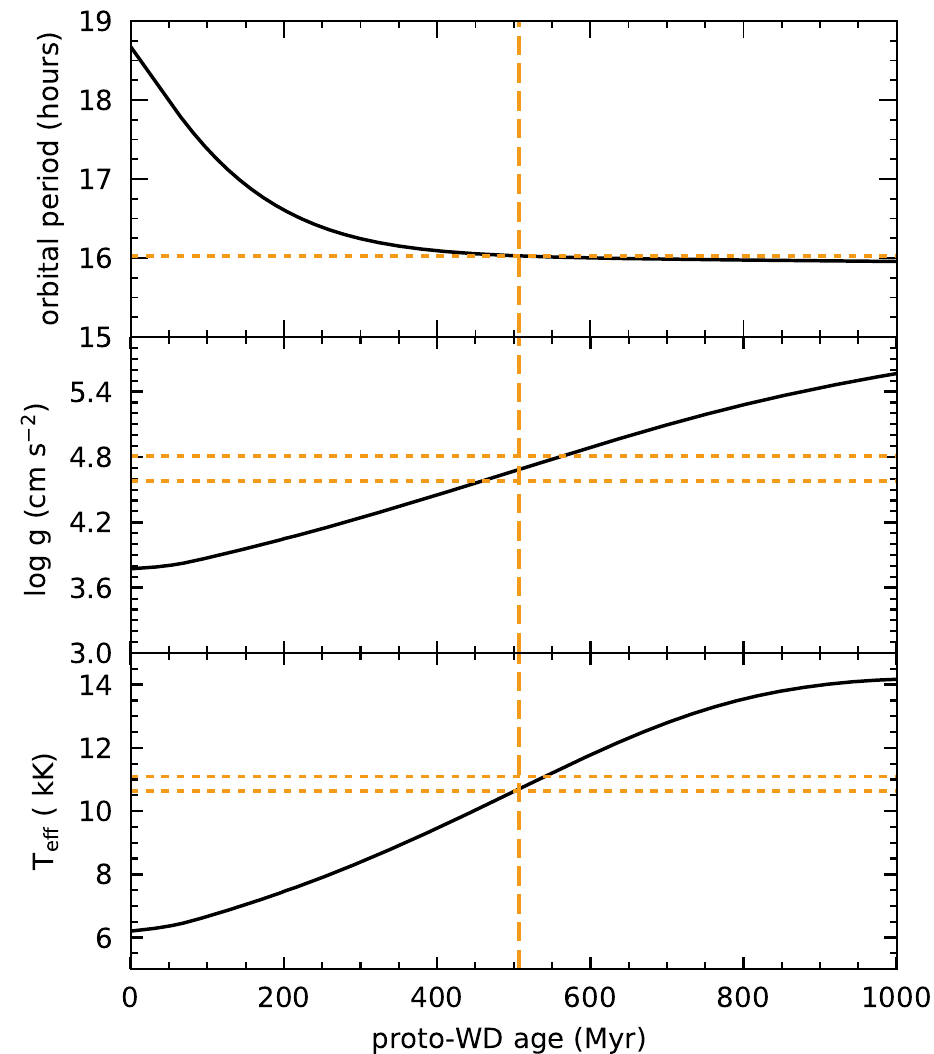}
\caption{Evolution of orbital period (top panel), surface gravity (middle panel) and effective temperature (bottom panel)  as function of the  proto-WD age. The horizontal orange dotted lines represent the observational constraints as given in Table~\ref{table:observed_parameters} while the vertical orange dashed line represents the time when the orbital period of the system matches the observed one.}
\label{fig:fig2}
\vspace{0.1in}
\end{figure}
%\vspace{1.0in}
\section{EVOLUTIONARY MODELS}

The evolutionary track followed by an ELM WD in the \teff-log $g$ plane, especially in the proto-WD phase, is very sensitive to the total mass and the assumed metallicity of the models. This important point is discussed thoroughly in \cite{istrate2016a}. Using the very precise estimates of the final masses of the A and B components of the
WASP 0247$-$25 system listed in Table~\ref{table:observed_parameters} as primary anchor points, we considered many different evolutionary sequences with varying metallicity and initial masses for the binary components. A first result is that we could not find, for any  reasonable metallicity, an evolutionary track for WASP 0247$-$25B in the narrow range of mass $M$(B) = 0.186 $\pm$ 0.002 \Msun~ that would go through the preferred location of  \cite{maxted2013} in the \teff-log $g$ diagram, i.e., \teff~= 11,380 $\pm$ 400 K and log $g$ = 4.576 $\pm$ 0.011\footnote{The location of WASP 0247$-$25B in Fig.~1 of \cite{istrate2016b} has been obtained by averaging the value of \teff~=~11,380 $\pm$ 400 K of Table~1 of  \cite{maxted2013} and that derived from   their preliminary analysis of the disentangled spectrum of the star  which leads to \teff~= 10,300 $\pm$ 200 K, while keeping the log $g$ value at 4.576 $\pm$ 0.011. The track going through that new location   shown in their figure show pulsational instabilities in the correct   range, but still corresponds to a mass of 0.189 $M_{\odot}$,  which is too large compared to the more stringent requirement of the present paper.} and  which satisfies in the same time the orbital constraints. A slightly  higher mass is needed, but we then encounter the problem of a mismatch between the predicted and the observed orbital period of the system at that point along the evolutionary track. Moreover, we also encounter difficulties in driving pulsation modes in view of the strong dependence of the blue edge on the total mass for a given metallicity, i.e., the more massive the star, the cooler the blue edge (see Figs.~2 and 3 of \citealt{istrate2016b} in this connection; also, \citealt{corsico2016}).

\begin{figure}
\includegraphics[width=1.05\columnwidth]{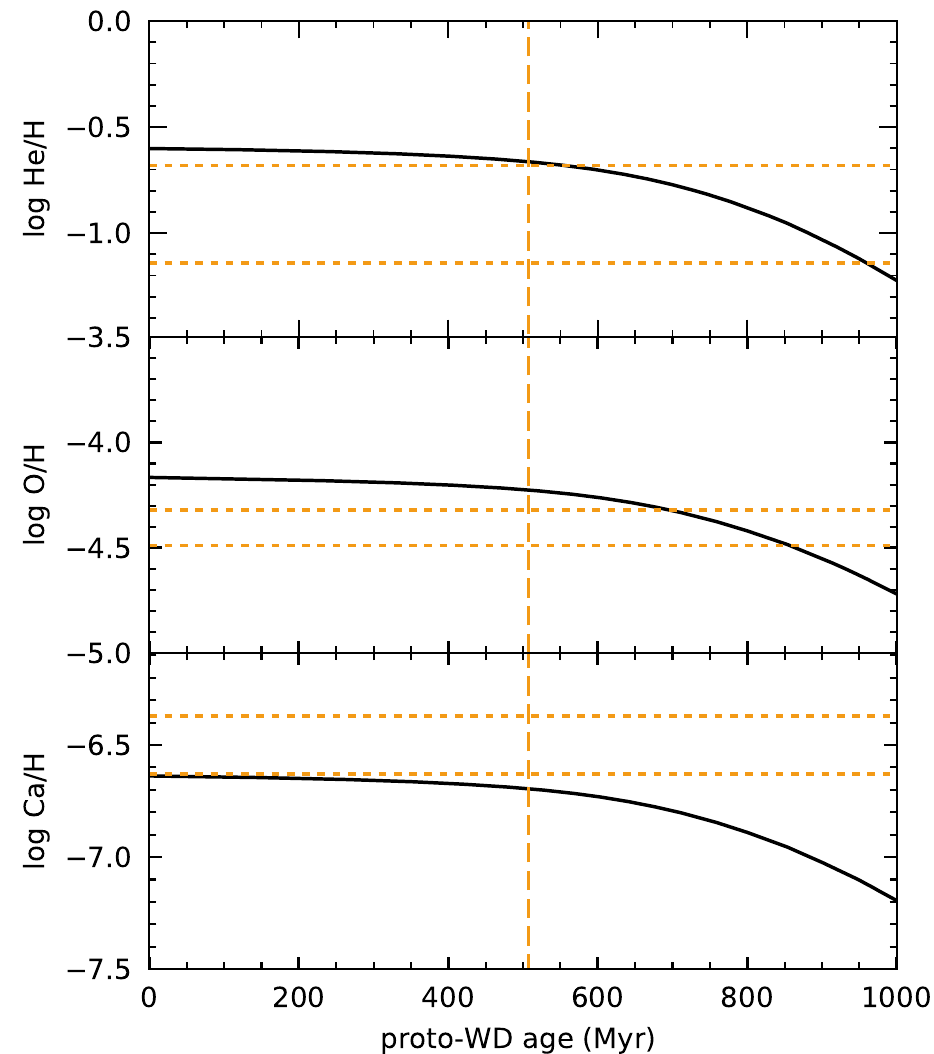}
\caption{  The evolution of surface abundances for  helium (top panel), oxygen (middle panel) and calcium (bottom panel) as function of the proto-WD age. The  orange dotted lines represent the observational constraints as given in Table~\ref{table:observed_parameters} while the  vertical orange dashed line represents the time when the orbital period matches the observed one.  }
\label{fig:fig3}
\end{figure}

On the other hand  we find a near-perfect solution (from an evolutionary point of view) for a 0.186 \Msun~track, computed with $Z$ = 0.0015 metallicity, that goes right through the location of WASP 0247$-$25B in the \teff-log $g$ diagram provided by the independent spectroscopic analysis of \citet{heuser2017}. The results are summarized in Figures  \ref{fig:fig1}, \ref{fig:fig2} and  \ref{fig:fig3}. Figure~\ref{fig:fig1} shows the evolutionary track in relation to the locations of WASP 0247$-$25B provided by \cite{maxted2013} and \cite{heuser2017} in the Kiel diagram. The top panel of Fig.~\ref{fig:fig2} illustrates that the orbital period of the system, which decreases (due to gravitational wave radiation) with increasing age, becomes equal to the observed orbital period after some 507 (507.42 to be more accurate) Myr of proto-WD evolution. We note that the final mass of
the A component in our  model system is 1.354~\Msun, well within the inferred value given in  Table \ref{table:observed_parameters}, i.e., $M$(A)~=~1.356~$\pm$~0.007~\Msun. The middle panel of Fig.~\ref{fig:fig2} illustrates that the surface gravity along the evolutionary track is log~$g$(B)~=~4.685 after 507~Myr of proto-WD evolution, in agreement with the
spectroscopic estimate of log~$g$(B)~=~4.70~$\pm$~0.11. Likewise, the bottom panel of Fig.~\ref{fig:fig2} shows the effective temperature as a function of age, indicates that it is equal to \teff(B)~=~10,700~K after 507~Myr of proto-WD evolution, in agreement with the spectroscopic estimate of \teff~= 10,870 $\pm$ 230 K.

Furthermore, Fig.~\ref{fig:fig3}  describes the  evolution of the surface composition of our model from the beginning of the proto-WD stage until an age of 1000~Myr. We find that the amount of He in the rotationally-mixed, homogeneous envelope of our model at 507~Myr proto-WD age, log~He/H(B)~=~$-$0.66, is just a tad larger than the 1-$\sigma$ upper limit of log~He/H(B)~=~$-$0.68 obtained in the spectroscopic analysis of \cite{heuser2017}. For its part, the predicted O abundance, log~O/H(B)~=~$-4.23$, is somewhat higher than the inferred spectroscopic value of $-4.40\pm^{0.08}_{0.09}$. In contrast, the  predicted Ca abundance, log~Ca/H(B)~=~$-$6.69, is slightly lower than the lower limit on the spectroscopic estimate given in Table~\ref{table:observed_parameters} (log~Ca/H(B)~=~$-$6.63). We note, in this context, that the relative metal composition that we used by default in our MESA calculations is that of \cite{gs1998}. Had we used instead the most recent evaluation of \cite{asplund09}, the predicted log~O/H(B) ratio would have been about 0.04 dex lower than found here, while the predicted log~Ca/H(B) value would have been some 0.09 dex higher, thus reducing the slight discrepancies depicted in Fig.~\ref{fig:fig3}.

The basic properties of our proposed evolutionary model at proto-WD age of 507 Myr are summarized in Table~\ref{table:model_parameters}. By construction, the orbital period is equal to the observed value of  \cite{maxted2013} and the initial masses of the components were picked to lead to final masses very similar to their measured values. In addition, the atmospheric properties of our model at that age are entirely consistent with the estimates obtained in the spectroscopic analysis of \cite{heuser2017}. Those, we recall, provide completely independent constraints from the inferences made in the light curve analysis of  \cite{maxted2013}. In this respect, we would like to point out that the differences we find with the preferred values of \cite{maxted2013} for \teff~and log $g$ remain small in an absolute sense.  Those differences probably imply some small systematic effects between the light curve analysis method of \cite{maxted2013} and the way our binary evolutionary models are calculated. Overall, however, we conclude that our assumed evolutionary model for WASP 0247$-$25B provides a rather realistic description of the basic properties of that star. Further on, we check if the model can also pass the test of seismology.

% Table 2
\begin{deluxetable}{cc}
\tablewidth{0pt}
%\rotate
\tabletypesize{\small}
\tablecaption{Parameters of the Model at Age 507 Myr}
\tablehead{
\colhead{Quantity} &
\colhead{Value} 
 }
\startdata
$M$(B) & 0.186 \Msun \\
$M$(A) & 1.354 \Msun \\
$P_{\rm orbit}$ & 16.027908 h \\
\teff(B)   & 10,700 K \\
log $g$(B) & 4.685 cm s$^{-2}$ \\
log He/H(B) & $-$0.66 \\
log Ca/H(B) & $-$6.69 \\
log O/H(B) & $-$4.23 \\
\enddata
\label{table:model_parameters}
\end{deluxetable}

\section{PULSATION CALCULATIONS}

\begin{figure}
\includegraphics[width=1.05\columnwidth]{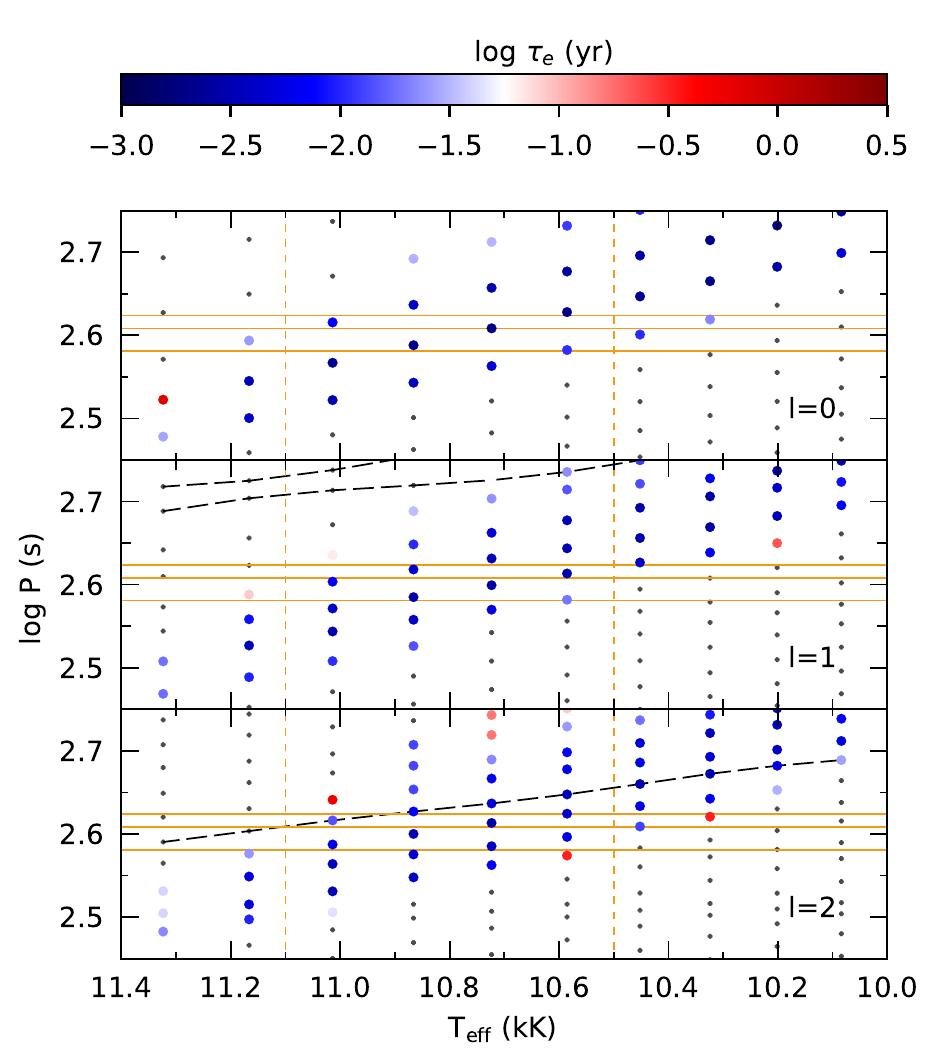}
\caption{Calculated period spectrum (small grey dots) along our evolutionary track for modes with degree index $\ell$ = 0 (top panel), $\ell$ = 1 (middle panel), and $\ell$ = 2 (bottom panel). The colored circles represent the excited modes color--coded according to the logarithm of the e-folding time. The dashed black curve in the bottom panel indicates the period of the fundamental mode ($n$ = 0) which divides the $p$-branch (below) from the $g$-branch (above) for the quadrupole modes. As is well known, such a fundamental mode does not
exist for dipole modes (for an isolated star), and the two dashed black curves indicate, respectively, the period of the lowest-order $p$-mode ($n$ = 1; lower curve) and that of the lowest-order $g$-mode ($n$ = $-$1; upper curve). The three detected periods in WASP 0247$-$25B are illustrated by the three horizontal orange lines. The uncertainties on these periods (see Table \ref{table:observed_parameters}) are too small to be seen at the scale of the present figure. The vertical dotted orange lines denote the region of interest in effective temperature (see text).}
\label{figure:period_spectrum}
\end{figure}

We carried out a nonadiabatic pulsation analysis of the individual equilibrium structures found along the retained evolutionary track described in the previous section. Figure \ref{figure:period_spectrum}  summarizes some of our results in the range of effective temperature of interest. Significantly, 
the figure indicates that the only acceptable seismic models are confined to the relatively narrow interval of 10,500 K $\lta$ \teff~ $\lta$ 11,100 K. On the cool side, the three detected periods are globally shorter than the periods found in the band of excited modes, while, on the warm side, the three observed periods are longer than those predicted unstable. Given that the effective temperature of our ``best'' evolutionary model at age 507 Myr described in Table \ref{table:model_parameters}  is well
within that interval, this constitutes a first consistency check with this important constraint coming out of our seismic analysis.

An examination of  Figure \ref{figure:period_spectrum} reveals that the pulsation periods are rather strongly dependent on the location of the equilibrium structures along the evolutionary proto-WD track (measured here by the
effective temperature). As the model evolves and gets hotter, the predicted periods (each defined by a given set of $\ell$-$n$ values) get shorter and some of them cross the range where the three observed periods are found. This implies that there must exist one or more values of the effective temperature corresponding to a best simultaneous match between the three detected periods and three of the periods belonging to the theoretical spectrum. In order to quantify this search for a structure providing the best fit to the observed periods, we use a standard $\chi^2$ approach. Hence, we look for minima of the quantity,

\begin{equation}
\chi^{2}=\sum_{i=1}^{3}\left(\frac{P_{{\rm obs}}^{(i)}-P_{{\rm
      th}}^{(i)}}{\sigma_{{\rm obs}}^{(i)}}\right)^{2}\quad,
\end{equation}

\noindent where the $P_{{\rm obs}}^{(i)}$'s and the ${\sigma_{{\rm obs}}^{(i)}}$'s are the three detected periods and their associated 1-$\sigma$ uncertainties as reported in Table~\ref{table:observed_parameters}, while the $P_{{\rm th}}^{(i)}$'s are the best-matching theoretical periods among the available values for a given location along the evolutionary track.\\
\begin{figure}[ht]
\includegraphics[width=\columnwidth]{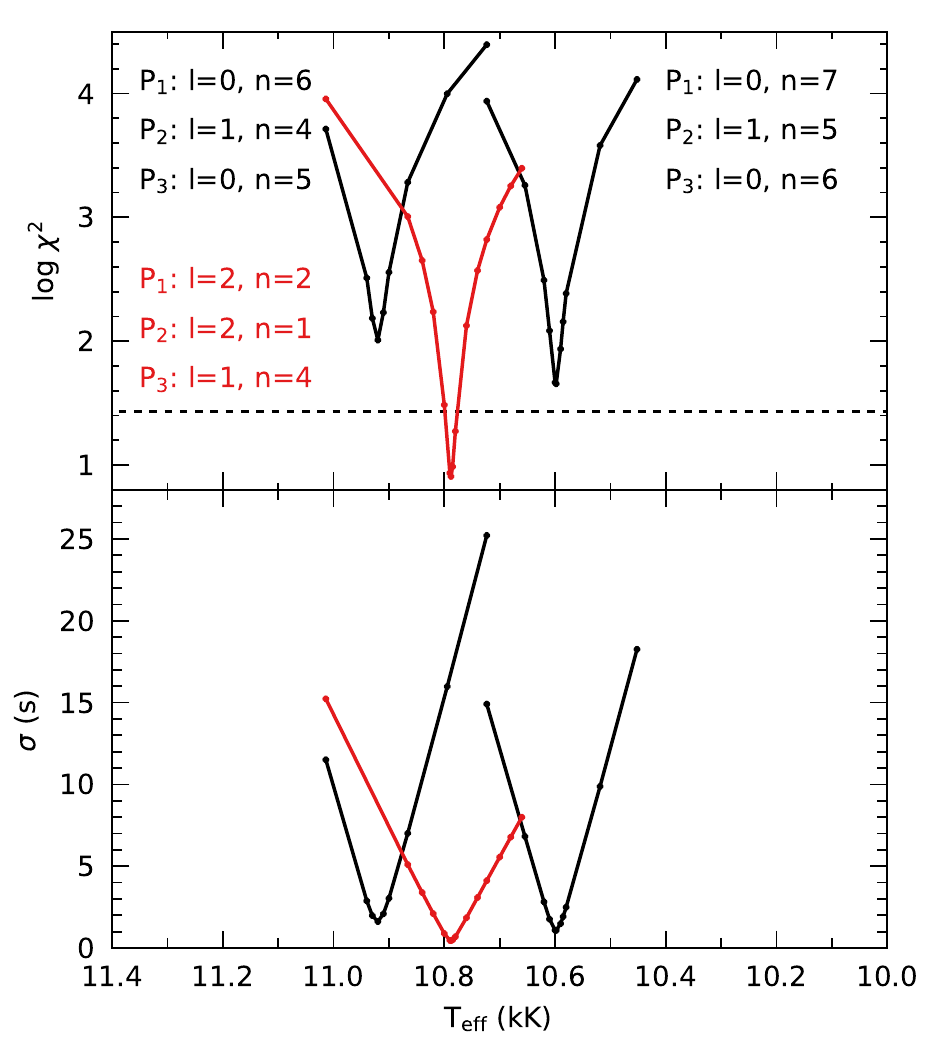}
\caption{Seismic solutions for the retained evolutionary track. The top panel illustrates the three possible solutions in terms of
well-defined minima in the formal $\chi^2$ along the track as measured by the effective temperature. If the three observed periods are restricted to the $\ell$ = 0 and/or 1 identification, then there are two possibilities of unequal quality as illustrated by the curves in black. Those differ in the mode identification by a change of $\Delta$ $n$ = $-$1 in the radial order of the modes, from the lower to the higher \teff. If, instead, modes with $\ell$ = 0, 1, or 2 are allowed, then the best-fit solution is given by the curve in red. The quality of the formal $\chi^2$ fit is estimated through the Q probability formalism; values of $\chi^2$ below the dotted horizontal line are considered good fits. The lower panel illustrates the variance $\sigma$ -- in units of second -- associated with each solution. }
\label{figure:seismic_solutions}
\end{figure}

If, for simplicity's sake, we assume that the detected pulsation modes belong to degree indices $\ell$ = 0 and/or 1, then we find two $\chi^2$ minima in the range of effective temperature of interest. A simple look at the top and middle panels of Figure \ref{figure:period_spectrum} is sufficient to anticipate such a minimum for an effective temperature just slightly higher than the value of \teff~= 10,585 K associated with one of the equilibrium structures along the evolutionary track. Likewise, corresponding to a change of $\Delta n$ = $-$1 for all three modes (but keeping the same identification for $\ell$), a second minimum can be anticipated for an effective temperature somewhat higher than \teff~= 10,866 K associated with another, more evolved, equilibrium structure. Also, clearly to the eye, this second minimum can only provide a lower quality fit than the previous one.

On the other hand, if we allow for the possibility that the detected modes belong to degree indices $\ell$ = 0, 1, and/or 2, then a quick examination of the middle and lower panels of Figure \ref{figure:period_spectrum} reveals that a potentially very good fit could be obtained for an effective temperature in between those of the equilibrium structures at 10,723 K and 10,866 K,
and closer to the former. Granted, the mode density is higher for the $\ell$ = 2 modes, so this increases the probability that a better period match is obtained by chance. Our standard $\chi^2$ approach is blind to that possibility, but we revisit this issue below. At the same time, the idea that quadrupole modes could be driven to observable amplitudes in a close binary system such as WASP 0247$-$25 has its own obvious appeal.

In summary, we find three possible seismic solutions in the sense of best matching the observed periods, while representing excited modes. In order to improve the ``resolution'' along the evolutionary track and pin down better the optimal period fit, we used parabolic interpolation in effective temperature. The results of that operation are shown in Figure \ref{figure:seismic_solutions} which shows, in the upper panel, the behavior of the least-squares coefficient $\chi^2$, and, in the lower panel, that of the variance $\sigma$ as a function of \teff~ for each of the three solutions. Taking into account the logarithmic scale used to depict $\chi^2$, each of the solutions corresponds to a very well-defined minimum.

This being said, a well-defined minimum in $\chi^2$ does not necessarily imply a good fit from a statistical point of view. Within the framework of our formal $\chi^2$ approach, it is possible to use the $Q$-probability formalism discussed in \cite{press1986} to estimate the quality of the fit (and see \citealt{randall2005} for an astrophysical application similar to the current one for more details). The usual adopted value for a $\chi^2$ fit to qualify as ``acceptable'' is $Q$ $\geq$ $1 \times 10^{-3}$, which corresponds in the present case with 3 degrees of freedom (0 free parameter and 3 data points) to $\chi^2$ $\leq$ 16.25. If we
relax this definition of an acceptable fit to include the solutions for which the three theoretical periods fall simultaneously within the $\pm$3-$\sigma$ ranges of the three detected periods, then the passing value is $Q$ $\geq$ $5.88 \times 10^{-6}$, which corresponds to $\chi^2$ $\leq$ 27.0. It is the latter threshold value that is plotted as a dotted horizontal line in the top panel of Figure \ref{figure:seismic_solutions}. The defining characteristics  of each of our three possible seismic solutions are listed in Table \ref{table3}. Details on the mode identification and period fits are provided in Table \ref{table4}.

\begin{deluxetable*}{ccccccccc}
\tablewidth{0pt}
%\rotate
\tabletypesize{\small}
\tablecaption{Defining Parameters for the Best-Fit Seismic Structures}
\tablehead{
\colhead{Solution} &
\colhead{$T_{\rm eff}$} &
\colhead{log $g$} &
\colhead{log He/H} &
\colhead{log O/H}  & 
\colhead{log Ca/H} &
\colhead{$\chi^2$} &
\colhead{$\sigma$} &
\colhead{$Q$} \\
\colhead{} &
\colhead{(K)} &
\colhead{(cm s$^{-2}$)} &
\colhead{} &
\colhead{} &
\colhead{} &
\colhead{} &
\colhead{(s)} &
\colhead{}
 }
\startdata
1 & 10598 & 4.666 & $-$0.66 & $-$4.22 & $-$6.69 & $4.53 \times 10^{1}$ & 1.08 & $8.07 \times
10^{-10}$ \\
2 & 10920 & 4.726 & $-$0.67 & $-$4.23 & $-$6.70 & $1.02 \times 10^{2}$ & 1.62 & $5.58 \times
10^{-22}$ \\
3 & 10788 & 4.701 & $-$0.67 & $-$4.23 & $-$6.69 & 8.09 & 0.46 & $4.42 \times
10^{-2}$ \\
\enddata

\label{table3}
\end{deluxetable*}

%\clearpage

% Table 4
\begin{deluxetable}{ccccccc}
\tablewidth{0pt}
%\rotate
\tabletypesize{\small}
\tablecaption{Mode Identification and Details of the Period Fits}
\tablehead{
\colhead{Solution} &
\colhead{$\ell$} &
\colhead{$n$} &
\colhead{$P_{\rm obs}$} &
\colhead{$P_{\rm th}$} &
\colhead{$\Delta P$} &
\colhead{$\Delta P/P$} \\
\colhead{} &
\colhead{} &
\colhead{} &
\colhead{(s)} &
\colhead{(s)} &
\colhead{(s)} &
\colhead{\%}
 }
\startdata
  & 0 & 7 & 380.95 $\pm$ 0.17 & 380.67 & $-$0.28 & 0.07 \\
1 & 1 & 5 & 405.82 $\pm$ 0.61 & 409.39 & 3.57    & 0.88 \\
  & 0 & 6 & 420.64 $\pm$ 0.76 & 422.82 & 2.18    & 0.52 \\ 
  &   &   &                   &        &         &      \\ 
  & 0 & 6 & 380.95 $\pm$ 0.17 & 380.51 & $-$0.44 & 0.12 \\
2 & 1 & 4 & 405.82 $\pm$ 0.61 & 410.24 & 4.42    & 1.09 \\
  & 0 & 5 & 420.64 $\pm$ 0.76 & 425.61 & 4.97    & 1.18 \\ 
  &   &   &                   &        &         &      \\
  & 2 & 2 & 380.95 $\pm$ 0.17 & 380.97 & 0.02    & 0.005 \\
3 & 2 & 1 & 405.82 $\pm$ 0.61 & 404.68 & $-$1.14 & 0.28 \\
  & 1 & 4 & 420.64 $\pm$ 0.76 & 422.26 & 1.62    & 0.39 \\
\enddata
\label{table4}
\end{deluxetable}

Formally speaking, only Solution 3 provides a statistically meaningful period match. That solution is also excellent by current seismic standards involving evolutionary models. Still, we think it unwise to dismiss the other two possibilities at this stage.
On the first account, Solution 3 proposes a mode identification that involves two $\ell=2$ modes out of the  three  observed ones, and this goes against the expected degree-amplitude correlation based on geometric cancellation effects \citep{dziembowski1977}. Note however, that sectorial $\ell=2$  modes could be favored in a close binary system such as WASP~0247-25B. Secondly, as indicated above, our $\chi^2$ method is insensitive to the question of mode density, i.e., if more modes are available in a given period window (as is the case considered in Solution 3 when the quadrupole modes are also included in the best-fit search), the chances of good matches naturally increase. To take into account that characteristic, \cite{brassard2001} proposed to introduce weights in a modified $\chi^2$ approach, which they used in their asteroseismic study of the pulsating
hot subdwarf B star PG 0014+067. The weights correspond to the inverse of the mode density for each family of modes with a given value of the degree index $\ell$ in a given period window. Translated into the present context, this means that such a modified $\chi^2$ for Solution 1 would improve by a factor of (24/15)$^2$ = 2.56 with respect to Solution 3 and perhaps become acceptable. These numbers are based on the observation that there are 7, 8, and 9 modes, respectively, with $\ell$ = 0, 1, and 2 in the observed period band in Figure \ref{figure:period_spectrum}. Thus, we submit that Solution 1 and Solution 2 should not be completely eliminated at this point, especially the former one. It is unfortunate, of course, that only three periods were detected in WASP 0247$-$25B as other pulsations would have provided much welcomed additional constraints on the possible seismic models that we can currently build.

In an ideal situation, our best evolutionary model described in Table \ref{table:model_parameters} would have shown three pulsation modes in its spectrum with periods nearly equal to the three detected periods in WASP 0247$-$25B. As it is, we found three possible seismic solutions (one of which qualitatively excellent) in the close vicinity of that best evolutionary model along the retained evolutionary track. The latter is defined by \teff~= 10700 K along that track, while, in comparison, seismic Solutions 1, 2, and 3 are found, respectively, at \teff~= 10598, 10920, and 10788 K. We note, in the present context, that the theoretical period spectra of proto-WD models are very sensitive to the total mass (see, e.g.,  \citealt{istrate2016b}), and this can be exploited to estimate what adjustment can be made to the total mass so that the best evolutionary model would coincide even better with a seismic solution than is the case here. For instance, on the basis of Solution 3 and using other sequences with
different masses, we estimate by interpolation that a model with a mass reduced by the small amount of 0.0009 \Msun~ from the current value of $M$(B) = 0.186 \Msun~ would show pulsation periods very close to the observed ones at the epoch along  its track when the predicted orbital period is equal to the measured one. We conclude from this discussion that our proposed model of WASP 0247$-$25B  (Table~\ref{table:model_parameters}) is consistent with the seismic information currently available.

\section{ROTATION PROPERTIES}

Our binary evolutionary models make specific predictions about the state of the internal rotation profile of the ELM proto-WD component. This can be exploited further to calculate the effects of rotational splitting on the pulsation spectrum in Fourier space. Given seismic observations of high enough temporal resolution (not yet available for WASP 0247$-$25B, however), this can be used to test further the proposed seismic solutions and, in particular, to check the derived mode identification and discriminate between the various possibilities.

While the  newly formed proto-WD is initially characterized by external layers rotating much faster than the inner regions (see  \citealt{istrate2016a} for details), differential rotation settles into solid-body rotation by the time a proto-WD has entered its final contraction phase toward its maximum effective temperature before turning into an ELM WD. 

Table \ref{table5} lists the solid-body rotation period, $P_{\rm rot}$, for each of the three seismic solutions, along with basic information concerning the mode identification associated with each solution. The predicted internal rotation period is very close to the measured orbital period at that stage in the evolution, and decreases with passing time (increasing \teff) as does the orbital period. 

\begin{deluxetable}{ccccccc}
\tablewidth{0pt}
%\rotate
\tabletypesize{\small}
\tablecaption{Rotation Properties}
\tablehead{
\colhead{Solution} &
\colhead{$\ell$} &
\colhead{$n$} &
\colhead{$P_{\rm th}$} &
\colhead{$P_{\rm rot}$} &
\colhead{$C_{\ell n}$} &
\colhead{$\Delta \nu_{\ell n}$} \\
\colhead{} &
\colhead{} &
\colhead{} &
\colhead{(s)} &
\colhead{(h)} &
\colhead{} &
\colhead{($\mu$Hz)}
 }
\startdata
  & 0 & 7 & 380.67 &        &         &  \\
1 & 1 & 5 & 409.39 & 16.034808 & 0.185    & 14.1 \\
  & 0 & 6 & 422.82 &        &         &  \\ 
  &   &   &                   &        &         &      \\ 
  & 0 & 6 & 380.51 &        &         &  \\
2 & 1 & 4 & 410.24 & 16.024922 & 0.214    & 13.6 \\
  & 0 & 5 & 425.61 &        &         &      \\ 
  &   &   &                   &        &         &      \\
  & 2 & 2 & 380.97 &        & 0.074    & 16.0 \\
3 & 2 & 1 & 404.68 & 16.027873 & 0.092 & 15.7 \\
  & 1 & 4 & 422.26 &        & 0.259    & 12.8 \\
\enddata
\label{table5}
\end{deluxetable}

For a given pulsation mode of frequency $\nu_{\ell n}$ (the inverse of the pulsation period), first-order perturbation theory predicts that solid-body rotation lifts the $2\ell + 1$ degeneracy associated with spherical symmetry and leads to a set of equally-spaced frequency components separated by,

\begin{equation}
\Delta \nu_{\ell n} = \frac {2\pi}{P_{\rm rot}}(1-C_{\ell n})\quad,
\end{equation}

\noindent where $C_{\ell n}$ is the first-order rigid rotation coefficient, which is specific to each degenerate mode specified by a set of indices $\ell$-$n$. That quantity can be computed from the unperturbed (rotation-free) eigenfunctions of the mode of interest and is reported in Table~\ref{table5}. As is well known, to first order, the pulsation frequency of a radial mode is not affected by rotation and there is also no possible rotational splitting in that case. On the other hand, dipole
modes split into triplets and quadrupole modes into quintuplets in Fourier space, with a spacing between adjacent frequency components given by $\Delta \nu_{\ell n}$. In the present case, the predicted spacings are given in the last column of Table~\ref{table5}. 

An important component in our ability to detect rotationally-split frequency components is the angle under which a given rotating
pulsator is observed. Within the framework of our binary evolutionary models, the spin axis of both components of the system should be perpendicular to the orbital plane. Under the reasonable assumption that the pulsation axis coincides with the rotation axis in WASP 0247$-$25B, the star should be seen nearly equator on since it is part of an eclipsing system. As a matter of fact, the inclination of the orbital plane of the system is $i = 87.3 \pm 0.9 \arcdeg$ according to \cite{maxted2013}. In that case, not all of the frequency components in a  rotationally-split multiplet are visible. For instance, Figure A5 of  \cite{charpinet2011} ( Supplementary Information) indicates that the central $m = 0$ component of an $\ell = 1$ triplet is invisible
when seen under an inclination angle of $90 \arcdeg$. Likewise, the $m = \pm 1$ components of an $\ell = 2$ quintuplet are also invisible when viewed under that angle. 

Referring back to Table~\ref{table5} , we thus find that, for both Solutions 1 and 2, the $\ell = 0$ pulsations at 381 s and 421 s should show up as singlets in the Fourier domain, while the $\ell = 1$ pulsation at 406 s should be seen as a doublet structure with components separated by $\Delta  \nu_{\ell n} = 2 \times 14.1~ \mu$Hz (Solution 1) or by $\Delta \nu_{\ell
 n} = 2 \times 13.6~ \mu$Hz (Solution 2). The mode identification is very different for Solution 3 and, in that case, the prediction is that the observed 381 s and 406 s pulsation -- identified as quadrupole modes -- should show triplet structures with adjacent components separated by $\Delta \nu_{\ell n} = 2 \times 16.0~ \mu$Hz (381 s) and $2 \times 15.7~ \mu$Hz (406 s). At the same time, the 421 s pulsation -- identified as a dipole mode -- should show a doublet structure with a spacing given by $\Delta
\nu_{\ell n} = 2 \times 12.8~ \mu$Hz. Of course, these predictions/diagnostics can only be of usefulness if and when
observations of sufficient temporal resolution become available in the future. This would imply a timebase covering several rotation/orbital periods. 

\section{CONCLUSION}

The primary goal of the present investigation has been to test our ability, using new evolutionary models including rotational mixing, in reproducing the properties of WASP 0247$-$25B, the best-studied pulsating ELM proto-WD so far \citep{maxted2013}. After investigating many possibilities, we isolated an evolutionary sequence featuring a final mass of 0.186 \Msun~ for the B component of the system (and a final mass of 1.354 \Msun~ for the A component) after mass transfer has ceased. Those values are, by full intent, perfectly compatible with the estimates listed in Table~\ref{table:observed_parameters}. Along that evolutionary track, we picked the structure corresponding to the epoch -- some 507 Myr after the beginning of the proto-WD phase -- when the orbital period of the system is exactly equal to the value reported in Table~\ref{table:observed_parameters}. Quite encouragingly,
this ``best'' evolutionary model of ours is found rather close to the location of WASP 0247$-$25B in the \teff-log $g$ diagram inferred by \cite{maxted2013}, although not rigorously at the same place. On the other hand, our best evolutionary model is located right within the 1-$\sigma$ error box in that diagram associated with the independent spectroscopic analysis of \cite{heuser2017}. Moreover, the predicted atmospheric abundances of He, O, and Ca in that evolutionary model are quite close to the spectroscopic values, although they are formally outside the estimated 1-$\sigma$ uncertainty ranges (see Table~\ref{table:model_parameters} in comparison to Table~\ref{table:observed_parameters}). 

On the pulsation front, we found that the only acceptable seismic models along the retained evolutionary track have to be within the relatively narrow interval 10,500 K $\lta$ \teff~$\lta$ 11,100 K, which encompasses, as it ideally should, our best evolutionary model at 10,700 K. This constraint comes from the requirement that the theoretical pulsation modes with periods falling within the range of the observed periods, 381$-$421 s, must be predicted excited in the model. In a perfect situation, the 10,700 K model would show three excited modes in its period spectrum that would match perfectly the three observed periods in  WASP 0247$-$25B. Instead, we found three possible seismic solutions in the close vicinity of our best evolutionary model along the
retained evolutionary track. Those are located at \teff~= 10,598 K (Solution 1), 10,920 K (Solution 2), and 10,788 K (Solution 3) along the track. Solutions 1 and 2 follow from the a priori assumption that the observed modes are either radial and/or dipole modes, while Solution 3 admits the presence of modes with either $\ell$ = 0, 1, and/or 2.

Solution 3 is definitely qualitatively and quantitatively superior and should be retained in principle, but, in a conservative spirit, we prefer to retain the other two possibilities at this stage. Given that our evolutionary approach also makes definite predictions about the state of internal rotation of WASP 0247$-$25B, we devised a test to identify the $\ell$ index of each observed mode through rotational splitting. As pointed out above, this will require follow-up observations with a temporal resolution significantly better than that achieved in the experiment carried out by \cite{maxted2013}. Given the eclipsing nature of the system, it will be hard work to improve on that past work, but the detection of rotational splitting would prove
ideal for discriminating between the three proposed seismic solutions. 

We note, in this context of mode identification, that the work of \cite{maxted2013} is also the source of additional information that might help in this regard. Indeed, from Supplementary Table 3 of \cite{maxted2013}, we can compute amplitude ratios such that, for instance, the ratio of the amplitude in the $u^\prime$ filter to that in the $r^\prime$ band is $A(u^\prime)/A(r^\prime)$ = 1.25 $\pm$ 0.08, 1.00 $\pm$ 0.40, and 1.64 $\pm$ 0.23 for the observed modes with periods of $P_1$, $P_2$, and $P_3$ (see Table 1), respectively. To interpret correctly these numbers, it will be necessary to develop the appropriate modeling
following, for example, \cite{randall2005}. This involves detailed work using specific intensities from proper model atmospheres, as well as nonadiabatic eigenfunctions. In the absence of such modeling for ELM proto-WDs, we can still make the point that ratios of the sort bear a unique signature of the $\ell$ index. For instance, in Solutions 1 and 2, both the $P_1$ and $P_3$ modes are identified as radial modes (see Table 4). Hence, the amplitude ratio should be exactly the same for both modes. To the extent that the value of $A(u^\prime)/A(r^\prime)$ = 1.25 $\pm$ 0.08 for the $P_1$ mode is truly significantly different from the value of $A(u^\prime)/A(r^\prime)$ = 1.64 $\pm$ 0.23 for the $P_3$ mode, this would immediately eliminate both seismic solutions. In contrast, Solution 3 boosts a different $\ell$ identification for mode $P_1$ compared to mode $P_3$, and this leads to different expected amplitude ratios as possibly observed. In truth, before the amplitude-color method can be exploited properly,
we need higher S/N observations to pin down better the pulsation amplitudes and their uncertainties in various filters. 

We thus plea for improved temporal resolution and higher S/N observations of the tantalizing eclipsing system WASP 0247$-$25. We
note, with some irony, that we are aware of one such an attempt with the SOAR telescope made in 2014 which, quite unfortunately, turned out unproductive. As it is, the pulsations seem to have ``disappeared'' below detectability on the one night when these follow-up observations were made (B. Dunlap 2017, private communication). Since it is difficult to imagine how the driving mechanism could be stopped over a timescale as short as a year or two, the more likely explanation could be amplitude modulation -- either intrinsic or due to beating between closely-spaced frequency components, the latter perhaps caused by rotational splitting -- that would cause the observable pulsation amplitudes to drop below detectability from time to time or periodically. Given that WASP 0247$-$25B has the potential for becoming the ``Rosetta Stone'' of ELM proto-WD seismology, we feel that further
efforts on the observational front would be much worthwhile. 

We conclude by pointing out that the comparison exercise that we carried out in this paper has turned out to be extremely encouraging in our quest for more realistic models of ELM proto-WDs. We indeed found it rather gratifying that the known properties of WASP 0247$-$25B could be closely reproduced  -- although, quite admittedly, not perfectly -- with our current evolutionary models that include rotational mixing as the main competitor against gravitational settling. Taking into account the
known constraints on the mass, orbital period, effective temperature, surface gravity, and atmospheric composition, there was no a priori guarantee that acceptable seismic solutions could be found for WASP 0247$-$25B. Yet, we found models along our retained evolutionary track that would both be compatible with these constraints and show pulsation modes that 1) have periods close to the observed values, and 2) are predicted excited. We hope that the required data become available some day, so that one can test our proposed seismic solutions either through detection of rotational splitting or through the application of the amplitude-color method.  It is likely that such follow-up observations would also reveal additional pulsation modes, which would help in refining our seismic models.

\acknowledgments
 We thank the anonymous referee for his/her positive feedback and constructive comments.
A.G.I. acknowledges support from the NASA Astrophysics Theory Program
through NASA grant NNX13AH43G. This work was also supported in part by
the NSERC Canada through a research grant awarded to G.F. The latter also
acknowledges the contribution of the Canada Research Chair
Program.

% References
\bibliographystyle{aasjournal}
\bibliography{wasp_bib}

\end{document}